\documentclass[aip, amsmath, amssymb,reprint]{revtex4-2}

\usepackage[utf8]{inputenc}
\usepackage{graphicx}
\usepackage{dcolumn}
\usepackage{bm}
\usepackage[utf8]{inputenc}
\usepackage[T1]{fontenc}
\usepackage{mathptmx}
\usepackage{gensymb}
\usepackage{upgreek}
\usepackage{xcolor}
\usepackage{textcomp}
\usepackage{hyperref}
\usepackage{mhchem}
\usepackage{float}
\usepackage[verbose]{placeins} 
\hypersetup{colorlinks=true,linkcolor=red,citecolor=blue,filecolor=magenta,urlcolor=magenta}

\usepackage{color,soul} 

\newcommand*{\citen}[1]{%
  \begingroup
    \romannumeral-`\x 
    \setcitestyle{numbers}%
    \cite{#1}%
  \endgroup   
}

\usepackage{titlesec}
\titleformat{\section}[hang]{\small\bfseries\sffamily}{\thesection.}{0.5em}{\MakeUppercase}
\titlespacing{\section}{0pc}{1pc}{0.2pc}

\begin{document}
\title{Squeezed Light Generation in Periodically Poled Thin-Film Lithium Niobate Waveguides}
\author{Xiaodong Shi}
\thanks{These authors contribute equally}
\affiliation{A$^\ast$STAR Quantum Innovation Centre (Q.InC), Agency for Science, Technology and Research (A$^\ast$STAR), 138634, Singapore}%
\affiliation{Institute of Materials Research and Engineering (IMRE), Agency for Science, Technology and Research (A$^\ast$STAR), 138634, Singapore}%

\author{Angela Anna Baiju}%
\thanks{These authors contribute equally}
\affiliation{A$^\ast$STAR Quantum Innovation Centre (Q.InC), Agency for Science, Technology and Research (A$^\ast$STAR), 138634, Singapore}%
\affiliation{Department of Physics, National University of Singapore, 117542, Singapore}
\affiliation{Centre for Quantum Technologies, National University of Singapore, 117543, Singapore}

\author{Xu Chen}%
\affiliation{Department of Materials Science and Engineering, National University of Singapore, 117575, Singapore}

\author{Sakthi Sanjeev Mohanraj}%
\affiliation{A$^\ast$STAR Quantum Innovation Centre (Q.InC), Agency for Science, Technology and Research (A$^\ast$STAR), 138634, Singapore}%
\affiliation{Institute of Materials Research and Engineering (IMRE), Agency for Science, Technology and Research (A$^\ast$STAR), 138634, Singapore}%
\affiliation{Department of Materials Science and Engineering, National University of Singapore, 117575, Singapore}

\author{Sihao Wang}%
\affiliation{A$^\ast$STAR Quantum Innovation Centre (Q.InC), Agency for Science, Technology and Research (A$^\ast$STAR), 138634, Singapore}%
\affiliation{Institute of Materials Research and Engineering (IMRE), Agency for Science, Technology and Research (A$^\ast$STAR), 138634, Singapore}%

\author{Veerendra Dhyani}%
\affiliation{A$^\ast$STAR Quantum Innovation Centre (Q.InC), Agency for Science, Technology and Research (A$^\ast$STAR), 138634, Singapore}%
\affiliation{Institute of Materials Research and Engineering (IMRE), Agency for Science, Technology and Research (A$^\ast$STAR), 138634, Singapore}%

\author{Biveen Shajilal}%
\affiliation{A$^\ast$STAR Quantum Innovation Centre (Q.InC), Agency for Science, Technology and Research (A$^\ast$STAR), 138634, Singapore}%
\affiliation{Institute of Materials Research and Engineering (IMRE), Agency for Science, Technology and Research (A$^\ast$STAR), 138634, Singapore}%

\author{Mengyao Zhao}%
\affiliation{Department of Materials Science and Engineering, National University of Singapore, 117575, Singapore}
\affiliation{Centre for Quantum Technologies, National University of Singapore, 117543, Singapore}

\author{Ran Yang}%
\affiliation{Department of Physics, National University of Singapore, 117542, Singapore}
\affiliation{Department of Materials Science and Engineering, National University of Singapore, 117575, Singapore}

\author{Yue Li}%
\affiliation{Department of Materials Science and Engineering, National University of Singapore, 117575, Singapore}

\author{Guangxing Wu}%
\affiliation{Department of Materials Science and Engineering, National University of Singapore, 117575, Singapore}
\affiliation{Centre for Quantum Technologies, National University of Singapore, 117543, Singapore}

\author{Hao Hao}%
\affiliation{Centre for Quantum Technologies, National University of Singapore, 117543, Singapore}

\author{Victor Leong}%
\affiliation{Institute of Materials Research and Engineering (IMRE), Agency for Science, Technology and Research (A$^\ast$STAR), 138634, Singapore}%

\author{Ping Koy Lam}%
\affiliation{A$^\ast$STAR Quantum Innovation Centre (Q.InC), Agency for Science, Technology and Research (A$^\ast$STAR), 138634, Singapore}%
\affiliation{Institute of Materials Research and Engineering (IMRE), Agency for Science, Technology and Research (A$^\ast$STAR), 138634, Singapore}%
\affiliation{Centre for Quantum Technologies, National University of Singapore, 117543, Singapore}

\author{Di Zhu}
\email{dizhu@nus.edu.sg}
\affiliation{Department of Materials Science and Engineering, National University of Singapore, 117575, Singapore}
\affiliation{Centre for Quantum Technologies, National University of Singapore, 117543, Singapore}
\affiliation{A$^\ast$STAR Quantum Innovation Centre (Q.InC), Agency for Science, Technology and Research (A$^\ast$STAR), 138634, Singapore}
\affiliation{Institute of Materials Research and Engineering (IMRE), Agency for Science, Technology and Research (A$^\ast$STAR), 138634, Singapore}


\begin{abstract}
Squeezed states of light play a key role in quantum-enhanced sensing and continuous-variable quantum information processing. 
Realizing integrated squeezed light sources is crucial for developing compact and scalable photonic quantum systems. 
In this work, we demonstrate on-chip broadband vacuum squeezing at telecommunication wavelengths on the thin-film lithium niobate (TFLN) platform. 
Our device integrates periodically poled lithium niobate (PPLN) nanophotonic waveguides with low-loss edge couplers, comprising bilayer inverse tapers and an SU-8 polymer waveguide. 
This configuration achieves a fiber-to-chip coupling loss of 1.4 dB and a total homodyne detection loss of 4 dB, enabling a measured squeezing level of 1.4 dB.
Additional measurements in a more efficient PPLN waveguide (without low-loss couplers) infer an on-chip squeezing level of approximately 10 dB at a pump power of 62 mW.
These results underscore the potential of TFLN platform for efficient and scalable squeezed light generation.
\end{abstract}
\maketitle

\section{Introduction}

Quadrature squeezed states of light, characterized by reduced quantum noise below the vacuum level along one quadrature component, are fundamental resources in continuous-variable (CV) quantum technologies, from quantum sensing and metrology to quantum computing and communications \cite{aasi2013enhanced,slusher2002squeezed,lawrie2019quantum,ourjoumtsev2006generating}.
Squeezed light can be generated in various physical platforms, such as atomic ensembles, optical fibers, and nonlinear crystals \cite{slusher1985observation,levenson1985generation,wu1986generation,slusher1987pulsed,andersen201630}. 
Among these, second-order ($\chi^{(2)}$) nonlinear crystals, such as periodically poled potassium titanyl phosphate (PPKTP) and periodically poled lithium niobate (PPLN), are most commonly adopted and have been consistently delivering high levels of squeezing based on spontaneous parametric down-conversion (SPDC) in various applications \cite{lam1999optimization,eberle2010quantum,vahlbruch2016detection,mehmet2011squeezed}.

The advent of integrated photonics has opened new avenues for realizing compact and scalable quantum photonic circuits. The tight optical confinement in nanophotonic waveguides substantially enhances nonlinear efficiency, reducing the pump power requirement for squeezing.
As a result, there has been a growing interest in developing on-chip squeezed light sources \cite{nehra2022few,williams2025ultrafast,shen2025strong,jia2025continuous,larsen2025integrated,liu2025wafer}.
However, translating the high squeezing levels achieved in bulk nonlinear systems into integrated platforms remains an outstanding challenge due to practical problems, such as waveguide propagation loss, coupling inefficiency, and limited power handling capacity \cite{o2009photonic,dutt2015chip}. 
Current efforts on integrated squeezers mainly focus on silicon nitride platforms via spontaneous four-wave mixing, a third-order ($\chi ^{(3)}$) nonlinear process \cite{larsen2025integrated,jia2025continuous,shen2025strong}. 
However, due to the relatively weak material-based $\chi ^{(3)}$ nonlinearity, optical cavities with high quality factors are needed. In addition, generating degenerate squeezing in such a system, essential for some quantum sensing and CV quantum computing applications, requires complicated cavity design or pumping schemes.

Recently, thin-film lithium niobate (TFLN) has emerged as a promising integrated $\chi^{(2)}$ nonlinear platform for quantum applications. 
Its strong $\chi^{(2)}$ nonlinearity, broad transparency window, low optical loss, and compatibility with quasi-phase matching via electrical poling make it well-suited for efficient nonlinear wavelength conversion and quantum light generation \cite{zhu2021integrated,javid2021ultrabroadband,javid2023chip,shi2024efficient,wang2018ultrahigh,kundu2024periodically}.
There have been a number of successful demonstrations of on-chip squeezers in quasi-phase-matched PPLN waveguides and cavities, as well as in modal-phase-matched TFLN microring resonators \cite{peace2022picosecond,park2024single,nehra2022few,williams2025ultrafast,arge2024demonstration,stokowski2023integrated,chen2022ultra}. 
Despite high nonlinearity, the measured squeezing levels reported on TFLN chips with homodyne detection are typically below 0.6 dB \cite{peace2022picosecond,park2024single,arge2024demonstration,stokowski2023integrated,chen2022ultra}.
This value is significantly lower than what has been achieved in bulk crystals, primarily due to high detection losses, where the coupling inefficiency between the TFLN waveguide and single-mode fiber or specific free-space mode plays a dominant role \cite{vahlbruch2008observation, mehmet2010observation}.

In this work, we directly measure 1.4 dB vacuum squeezing from a single-pass, periodically poled, TFLN waveguide with continuous-wave pump and homodyne detection. We integrate a high-quality PPLN nanophotonic waveguide with an efficient chip-to-fiber coupler, consisting of bi-layer inverse tapers and an SU-8 polymer waveguide, enabling a low coupling loss of 1.4 dB into a lensed telecom single-mode fiber.
The low-loss coupler allows us to reduce the total homodyne detection loss down to 4 dB, leading to a measured broadband squeezing of 1.4 dB ($\sim$4.7 dB on-chip squeezing) at 38 mW on-chip pump power.
We also experimentally demonstrate a more efficient PPLN waveguide (without low-loss couplers), and infer an on-chip squeezing level of $\sim$10 dB at a pump power of 62 mW.
These results underscore the potential of TFLN platform in efficient and scalable squeezed light generation.

\section{Methods and Results}
 
The core component of our squeezer is a PPLN nanophotonic waveguide.
We design the waveguide based on a 600 nm thick x-cut TFLN with an etch depth of 300 nm and a width of 2 \textmu m. 
Type-0 phase matching between fundamental transverse-electric (TE) modes at 780 nm and 1560 nm can be achieved with a poling period of 4.76 \textmu m, utilizing the maximal $d_{33}$ nonlinear coefficient. 
The fabrication begins with patterning the poling electrodes on the TFLN (NanoLN) using electron-beam (e-beam) lithography, followed by metal evaporation and lift-off. 
Poling is then performed by applying a sequence of high-voltage pulses to induce periodic domain inversion. 
After periodic poling, we fabricate the waveguide within the poled region using e-beam lithography, followed by dry etching using inductively coupled plasma reactive ion etching (ICP-RIE) and silicon dioxide cladding using plasma-enhanced chemical vapor deposition (PECVD).

\begin{figure}[htbp]
\center 
\includegraphics[width=0.99\linewidth]{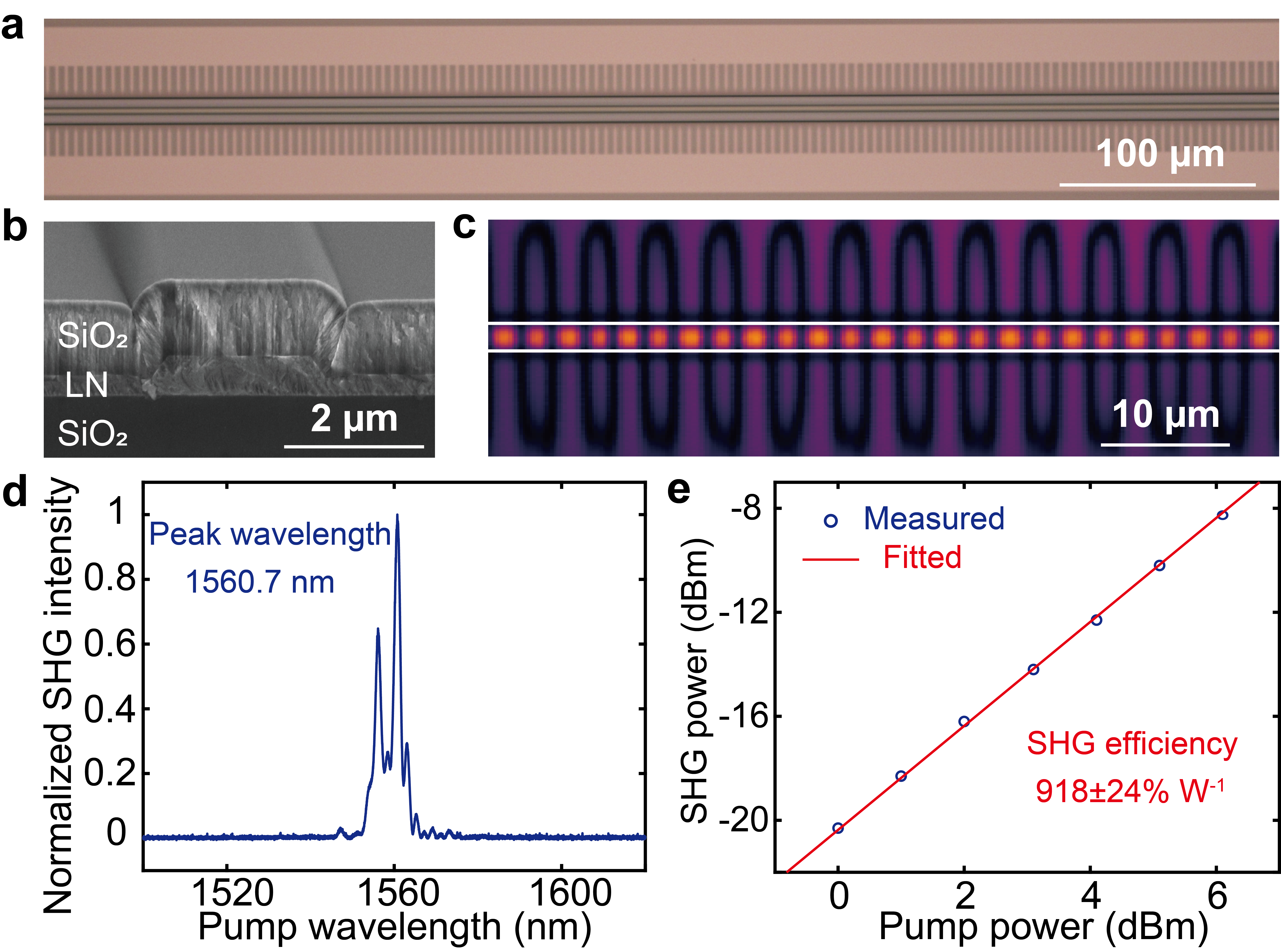}
\caption{\textbf{Fabrication and characterization of periodically poled lithium niobate (PPLN) nanophotonic waveguides.} \textbf{a,} Top-view optical micrograph, \textbf{b,} cross-section SEM, and \textbf{c,} top-view laser-scanning SHG microscopy image of a fabricated PPLN nanophotonic waveguide. \textbf{d,} Measured SHG spectrum from the PPLN waveguide, where the main peak is at 1560.7 nm. \textbf{e,} On-chip SHG power as a function of on-chip pump power in the 1 cm long PPLN waveguide. A linear fitting reveals an on-chip SHG efficiency of 918$\pm$24\% W$^{-1}$.}
\label{Fig1}
\end{figure}

Figure \ref{Fig1}\textbf{a} and \ref{Fig1}\textbf{b} show a top-view optical micrograph and a cross-section scanning electron micrograph (SEM) of the fabricated PPLN nanophotonic waveguide, respectively.
Figure \ref{Fig1}\textbf{c} presents a top-view second-harmonic laser scanning micrograph of the PPLN waveguide, indicating periodic domain inversions along the waveguide. 

To determine the waveguide loss, we measure a microring resonator fabricated on the same chip. From its intrinsic quality factors of $\sim$1.8$\times$10$^6$ around 1560 nm, we extract a propagation loss of $\sim$0.2 dB/cm.
To characterize the nonlinear performance of the PPLN waveguide, we measure its second harmonic generation (SHG) response.
Figure \ref{Fig1}\textbf{d} shows its SHG spectrum by sweeping a tunable laser while recording the second harmonic power at the output.
The primary peak appears at a pump wavelength of 1560.7 nm (SHG wavelength at 780.35 nm).
The side peaks are likely due to phase-matching shifts induced by the thickness non-uniformity of the LN thin film.
We also characterize the SHG power at different pump powers (Fig. \ref{Fig1}\textbf{e}).
The linear fitting reveals an on-chip SHG efficiency of 918 $\pm$ 24 \% W$^{-1}$ in the PPLN waveguide.

\begin{figure*}[htbp]
\center 
\includegraphics[width=0.8\linewidth]{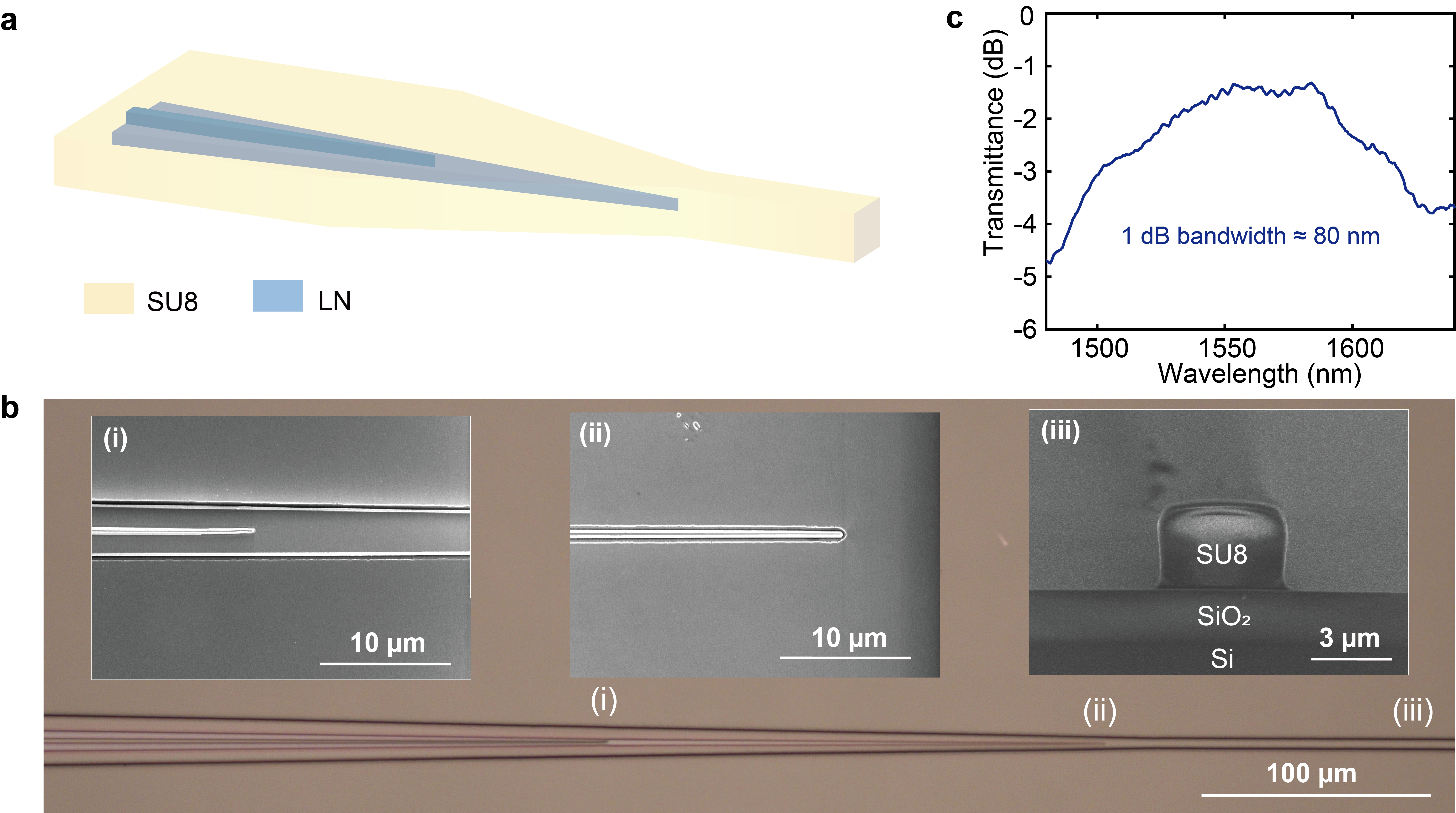}
\caption{\textbf{Efficient and broadband chip-to-fiber coupling on thin-film lithium niobate (TFLN) integrated platform.} \textbf{a,} A schematic of the coupling scheme in 600 nm thick TFLN. Light couples from a 600 nm thick LN rib waveguide to 300 nm thick LN ridge waveguide through an inverse taper in the top 300 nm thick LN layer, and evanescently couples to the SU-8 waveguide through a second inverse taper in the bottom 300 nm thick LN layer. \textbf{b,} Micrograph of the fabricated chip-to-fiber coupler. Zoomed-in SEMs of the inverse tapers in the inset (i) top and inset (ii) bottom LN layers after etching. Inset (iii) Cross-section SEM of the SU-8 edge coupler. \textbf{c,} Measured transmission spectrum of a chip-to-fiber coupler, showing a coupling loss of 1.4 dB around 1560 nm with a 1 dB transmission bandwidth of 80 nm.}
\label{Fig2}
\end{figure*}

Efficient chip-to-fiber coupling loss plays a vital role in squeezing measurement.
To minimize coupling loss, we implement a low-loss edge coupler at the output of the PPLN waveguide (Fig.~\ref{Fig2}\textbf{a}).
It consists of an inverse taper in the 600 nm thick LN rib waveguide to couple the light to the 300 nm thick LN ridge waveguide, and a second inverse taper in the ridge waveguide to couple the light to the SU-8 polymer waveguide on top, terminating at the edge of the chip \cite{ying2021low,he2019low}.
It has a dimension of 3.4$\times$3.2 \textmu m$^2$, enabling better mode matching between the waveguide facet and a lensed fiber (spot diameter of $\sim$3 \textmu m).

In the fabrication process, after patterning the PPLN waveguide with first-layer inverse taper, we use a second e-beam lithography and dry etching process to define the slab waveguide and the second inverse taper.
After PECVD cladding, to expose the coupler region, we open a window in the cladding via ultra-violet (UV) lithography and hydrochloric acid (HF) wet etching.
Finally, the SU-8 waveguide is patterned through another UV lithography step.

Figure \ref{Fig2}\textbf{b} shows the fabricated mode converter for low-loss chip-to-fiber coupling.
Insets (i) and (ii) show zoomed-in SEMs of the inverse tapers in the top and bottom LN layers after two-step etching, respectively.
Inset (iii) shows the cross-section SEM of an exposed SU-8 waveguide end facet after cleaving.
To quantify the coupling loss, we compare two devices fabricated on the same chip: (i) the PPLN waveguide with no mode converter at the input but a low-loss coupler (double-layer taper with SU8 waveguide) at the output, and (ii) a reference waveguide with no mode converters at input or output. The total fiber-to-fiber transmission around 1560 nm is measured to be 6.4 dB for the PPLN device and 9.8 dB for the reference waveguide. Since both devices share the same input coupler design and waveguide geometry, this allows us to infer the improvement from the low-loss output coupler. 
From this comparison, we estimate the input coupling loss of the 2 \textmu m-wide waveguide to be 4.8 dB/facet and the output coupling loss of the mode converter to be 1.4 dB/facet around 1560 nm, taking into account the 0.2 dB propagation loss.
Figure \ref{Fig2}\textbf{c} shows a measured transmission spectrum of the edge coupler, exhibiting a 1 dB transmission bandwidth of $\sim$80 nm, from 1520 to 1600 nm.

\begin{figure*}[htbp]
\center 
\includegraphics[width=0.8\linewidth]{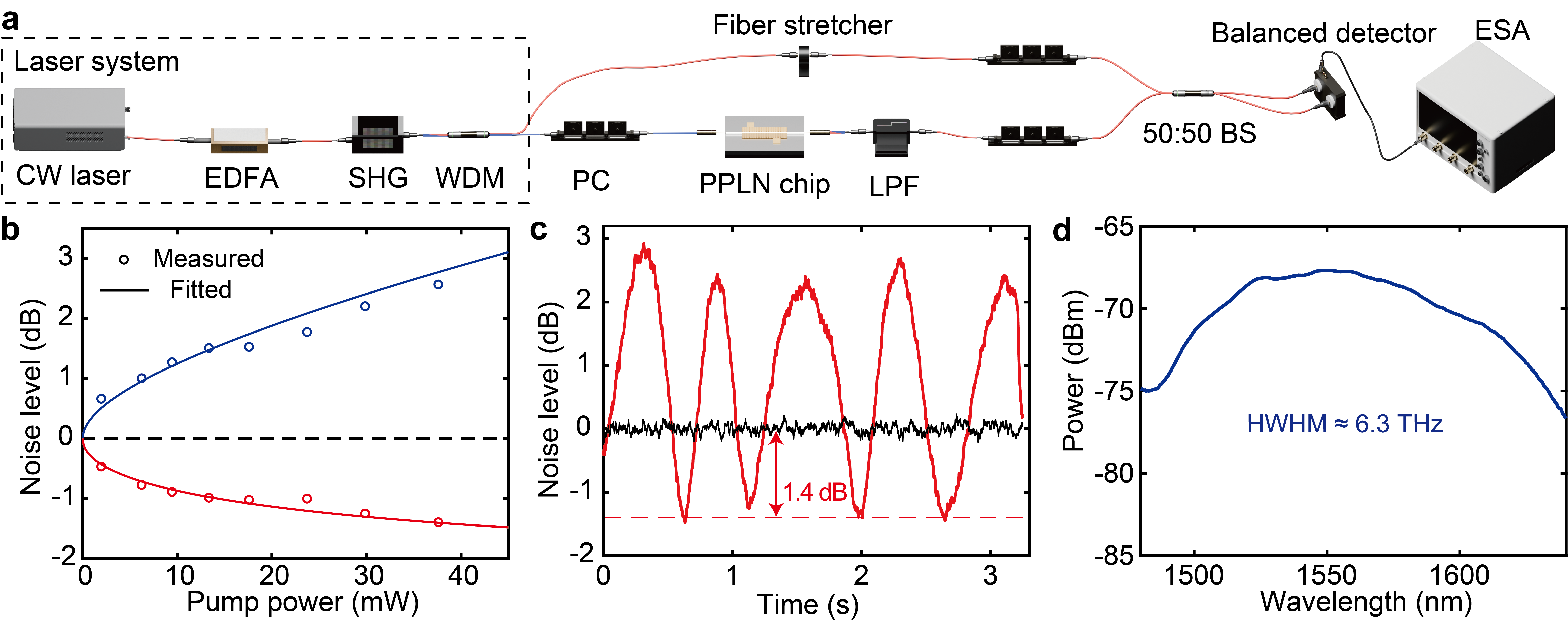}
\caption{\textbf{Squeezed light generation in a PPLN nanophotonic waveguide.} \textbf{a,} Measurement setup. CW laser: continuous-wave laser, EDFA: erbium-doped fiber amplifier, SHG: second-harmonic generation module, WDM: wavelength division multiplexer, PC: polarization controller, LPF: long-pass filter, BS: beam splitter, balanced detector, ESA: electrical spectrum analyzer. \textbf{b,} Measured and simulated squeezing and anti-squeezing levels as a function of pump power. The errors in noise fluctuations are on the order of 0.1 dB, and therefore, not marked on the plot. \textbf{c,} Normalized noise power (red line) at a pump power of 38 mW, while the LO phase is being tuned continuously. Black line shows the shot-noise level. \textbf{d,} Squeezed light spectrum showing a half-width at half-maximum (HWHM) of $\sim$6.3 THz.}
\label{Fig3}
\end{figure*}

We then measure squeezed light generation in the above PPLN nanophotonic waveguide with chip-to-fiber coupler.
Figure \ref{Fig3}\textbf{a} shows the schematic of the experimental setup. 
The customized laser system (Precilasers) consists of a continuous-wave (CW) seed laser at fundamental-harmonic (FH) wavelength of 1560.7 nm, amplified by an erbium-doped fiber amplifier (EDFA) and then directed to a SHG module to generate second-harmonic (SH) light at 780.35 nm.
The undepleted FH pump light is separated out from the SH light and used as the local oscillator (LO). 
The LO passes through a polarization controller (PC) and a piezo-actuated fiber stretcher controlled by an arbitrary waveform generator for polarization control and continuous phase tuning, respectively. 
Meanwhile, SH light, after passing through a PC, is coupled to the PPLN waveguide (Fig. \ref{Fig1}\textbf{a}) via a lensed fiber. 
A temperature controller is used to stabilize the phase-matching wavelength of the PPLN waveguide at 1560.7 nm. 
The generated squeezed light is collected from the fiber-to-chip coupler (Fig. \ref{Fig2}\textbf{b}) using a second single-mode telecom lensed fiber. 
After pump filtering, this path is combined with the LO at a 50:50 beam splitter (BS) for homodyne detection using a balanced detector (Wieserlabs WL-BPD1GA). 
The detector output is sent to an electronic spectrum analyzer (ESA; Agilent Technologies N9000A) for noise characterization.
The ESA measurements are taken at a central frequency of 100 MHz, resolution bandwidth (RBW) of 100 kHz, and video bandwidth (VBW) of 10 Hz.
Figure \ref{Fig3}\textbf{b} shows the measured squeezing (red circles) and anti-squeezing (blue circles) levels as a function of on-chip pump power in the PPLN waveguide.
We observe 1.4 $\pm$ 0.1 dB squeezing and 2.6 $\pm$ 0.2 dB anti-squeezing at a pump power of 38 mW.
Figure \ref{Fig3}\textbf{c} shows the corresponding normalized noise power as a function of time with continuous phase tuning. 
The reported squeezing and anti-squeezing values correspond to the mean of measured squeezing troughs and anti-squeezing peaks, respectively, while the uncertainties are the standard deviations.
The detection loss after the PPLN waveguide is estimated to be 3.8 dB, including chip-to-fiber coupling loss of 1.4 dB, pump filter insertion loss of 1.1 dB, 50:50 BS insertion loss of 0.3 dB (each output port shows $\sim$-3.3 dB attenuation relative to the input), and detector loss of 1.0 dB.
Taking all the detection losses into account, we infer a limited on-chip squeezing of $\sim$4.7 dB at a pump power of 38 mW, due to 0.2 dB intrinsic PPLN waveguide loss.
Since we perform continuous scan of the LO phase, the effect of anti-squeezing quadrature coupling to the squeezing quadrature due to LO phase noise can be neglected.
As our system is loss-limited, the measured values can be theoretically approximated using \cite{bachor2019guide,nehra2022few,chen2022ultra,eto2008observation} 
\begin{equation}
    S_{\pm} = 10\log_{10}(1-T+T e^{\pm 2 \sqrt{\eta P}}),
    \label{eq1}
\end{equation}
where $S_{-}$ and $S_{+}$ indicate the squeezing and anti-squeezing levels in decibel scale, respectively. $T$ is the total efficiency (including PPLN waveguide loss and detection losses), $\eta$ is the SHG efficiency, and $P$ is the pump power.
The calculated results of on-chip squeezing (red line) and anti-squeezing (blue line) levels are shown in Fig. \ref{Fig3}\textbf{b}, matching well with the measurement results.
The equation also indicates that increasing the pump power can further enhance the squeezing levels. However, in our experiments we find that at on-chip pump powers exceeding $\sim$100 mW, local heating at the LN inverse taper could cause damage, likely due to the high optical intensity in the narrowing taper region. To address this limitation, future work could explore alternative coupler geometries such as trident edge couplers, which expand the mode area gradually and can tolerate higher pump powers \cite{liang2022efficient}.
We also measure the SPDC spectrum from this PPLN waveguide (Fig. \ref{Fig3}\textbf{d}) by coupling one of the BS outputs to an optical spectrum analyzer (OSA).
It shows a broadband squeezed light generation with a half-width at half-maximum (HWHM) of $\sim$6.3 THz.

\section{Discussion and Conclusion}
 
\begin{figure}[tbph]
\center 
\includegraphics[width=0.99\linewidth]{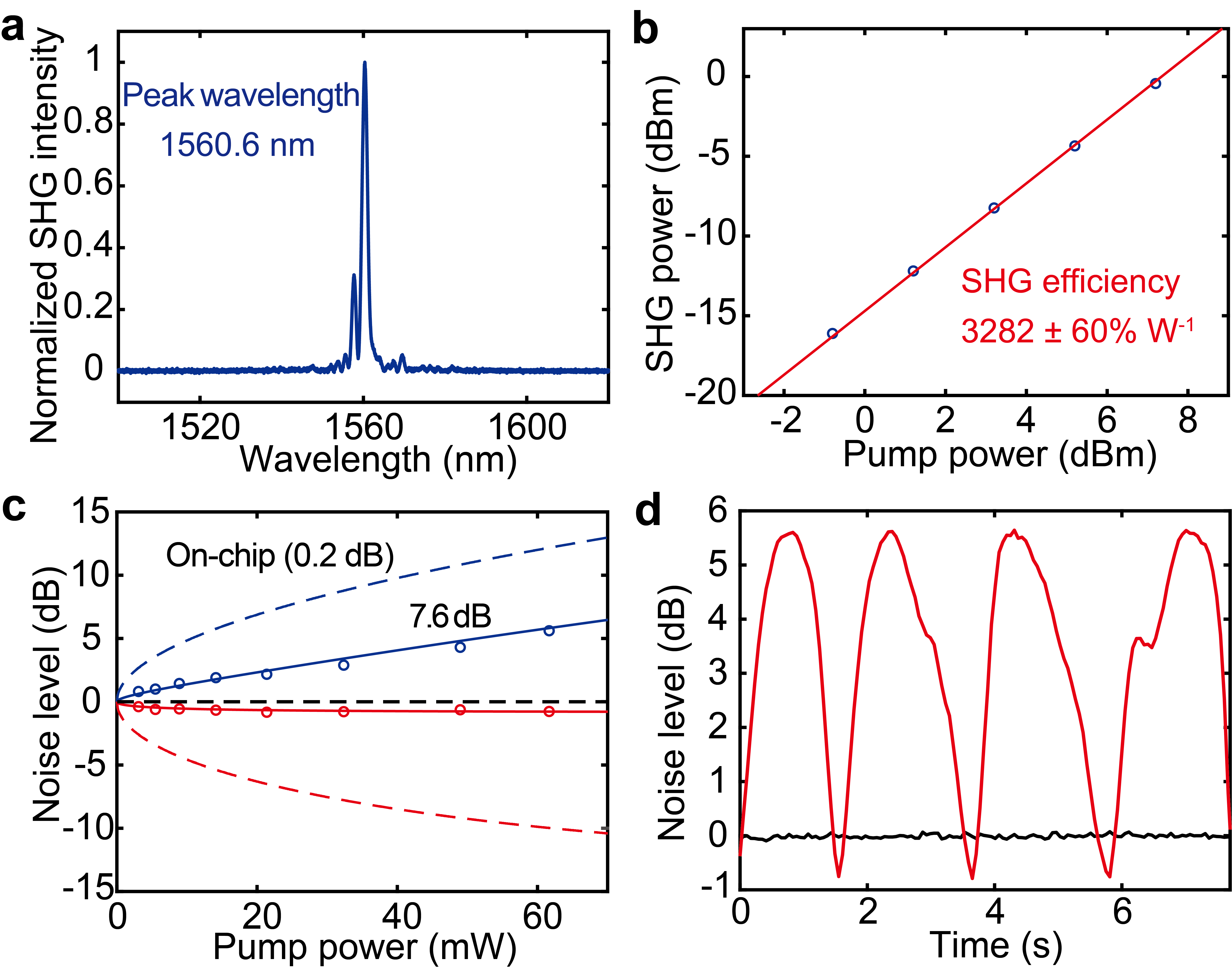}
\caption{\textbf{10 dB on-chip squeezing in an efficient PPLN nanophotonic waveguide.} \textbf{a,} Measured SHG spectrum from the 1.1 cm long PPLN waveguide, with a sharp peak at a pump wavelength of 1560.6 nm. \textbf{b,} SHG power as a function of pump power, showing a high SHG efficiency of 3282 $\pm$ 60 \% W$^{-1}$. \textbf{c,} Simulated and measured squeezing and anti-squeezing levels as a function of pump power, for different assumed losses, indicating an on-chip squeezing of $\sim$10 dB. \textbf{d,}  Normalized noise power as a function of time with phase tuning at a pump power of 68 mW.}
\label{Fig4}
\end{figure}

The on-chip squeezing level largely depends on the nonlinear efficiency of the waveguide.
We fabricate a more efficient 1.1 cm long PPLN waveguide (without low-loss coupler).
The major improvement includes better poling quality (duty cycle control) and the use of a TFLN chip with improved thickness uniformity, which reduces phase-matching wavelength fluctuations along the waveguide. 
This waveguide shows a sharp phase-matching peak at 1560.6 nm (Fig. \ref{Fig4}\textbf{a}) and a high on-chip SHG efficiency of 3282 $\pm$ 60 \% W$^{-1}$ (Fig. \ref{Fig4}\textbf{b}).

Figure \ref{Fig4}\textbf{c} shows the measured squeezing (red circles) and anti-squeezing (blue circles) levels as a function of on-chip pump power in this waveguide.
We observe 0.77 $\pm$ 0.02 dB squeezing 5.62 $\pm$ 0.02 dB anti-squeezing at a pump power of 62 mW. 
Figure \ref{Fig4}\textbf{d} shows the corresponding normalized noise power as a function of time with continuous phase tuning.
Although the nonlinear efficiency of this waveguide is high, without the integrated mode converter for efficient chip-to-fiber coupling, the detection loss after the PPLN waveguide is also high ($\sim$7.4 dB), resulting in a low measured squeezing level.
On the other hand, the anti-squeezing is less affected by loss.
Accounting for the detection loss, our results suggest an inferred on-chip squeezing of $\sim$10 dB at a pump power of 62 mW.
The measured squeezing level can be improved substantially by further optimizing the chip-to-fiber coupling and optical component losses. 
Without the intrinsic PPLN waveguide loss of $\sim$0.2 dB, the squeezing generated in a lossless PPLN waveguide would be $\sim$12 dB.

\begin{table*}[htbhp]
\caption{\label{table1} A summary of reported squeezers on TFLN platform, including device structure, pump scheme, detection scheme, waveguide length ($L$), SHG efficiency ($\eta$), total loss ($Loss$), on-chip pump power (peak power for pulsed pump) ($P$), and measured squeezing level ($S_{-}$).}
\begin{tabular}{ w{c}{1.6cm} w{c}{1.8cm}  w{c}{1.6cm}  w{c}{2.6cm} w{c}{1.4cm} w{c}{1.5cm} w{c}{1.4cm}  w{c}{1.4cm}  w{c}{1.4cm}}
 Ref.&Structure&Pump scheme& Detection scheme& $L$& $\eta$& $Loss$ & $P$ &$S_{-}$\\
  & &&&(cm)&(\%W$^{-1}$)& (dB) & (mW) & (dB)\\
  \hline
[\citen{nehra2022few}]&Waveguide& fs pulse& OPA + OSA& 0.25& -& 1.5& $\sim$10$^4$ &4.9\\

[\citen{williams2025ultrafast}]&Waveguide& fs pulse&OPA + OSA& 0.25& -&2.6& $\sim$10$^5$&4.1\\

[\citen{peace2022picosecond}]&Waveguide& ps pulse &Homodyne& 0.47& 28 &6.6&<300&0.33\\

[\citen{park2024single}]&Cavity&CW &Homodyne& -& -& 7.4&4&0.55\\

[\citen{arge2024demonstration}]&Cavity& CW&Homodyne& -& -&6.2&6.9&0.46\\

[\citen{stokowski2023integrated}]&Waveguide&CW &OPA + Homodyne& 1& 1000&7&3.4&0.12\\

[\citen{chen2022ultra}]&Waveguide&CW  &Homodyne& 0.5& 155&7.1&82&0.56\\

This work &Waveguide& CW &Homodyne& 1.1 &3282&7.6&62&0.77\\
This work &Waveguide& CW &Homodyne& 1.0& 918 &4.0&38&1.4\\
\end{tabular}
\end{table*}

Table \ref{table1} summarizes squeezers that have been demonstrated in the TFLN integrated platform.
Compared to those using homodyne detection, our results show the highest measured squeezing level, which benefits from the relatively high nonlinear efficiency and lower detection losses.
With further performance improvements and future integration with other functional components, we expect TFLN to become a strong candidate in realizing scalable and complex photonic circuits for practical quantum  applications in CV quantum computing, simulation, and sensing.

\begin{acknowledgments}
This research is supported by the National Research Foundation Singapore (NRF-NRFF15-2023-0005), A*STAR (M23M7c0125), and Centre for Quantum Technologies Funding Initiative (S24Q2d0009).
\end{acknowledgments}

\section*{Data Availability Statement}

The data that support the findings of this study are available from the corresponding author upon reasonable request.

\section*{REFERENCES}

\bibliography{References}

\end{document}